# EASY JAVA SIMULATION, INNOVATIVE TOOL FOR TEACHERS AS DESIGNERS OF GRAVITY-PHYSICS COMPUTER MODELS

**Loo Kang WEE,** [1]*Ministry of Education, Educational Technology Division, Singapore*
**Giam Hwee GOH,** [2]*Ministry of Education, Yishun Junior College, Singapore*
**Ee-Peow LIM,** [3]*Ministry of Education, Anderson Junior College, Singapore*

## Abstract
This paper is on the customization of computer models using the Easy Java Simulation Authoring Toolkit for the Singapore syllabus, based on real data, supported with literature-reviewed pedagogical features. These four new computer models serve to support the enactment of evidence-based classroom science learning that is. The authors suggest that with strong teacher guidance, these computer models can be effective tools for science learning. Students also enjoy learning because of the increased level of interactivity when exploring the modeled science with enhanced visualization as compared with the traditional pen and paper problem solving.
Pilot research suggests that students' enactment of investigative learning like scientists is now possible where gravity-physics 'comes alive' and students' learning is enriched.
Download simulations:
https://dl.dropboxusercontent.com/u/44365627/lookangEJSworkspace/export/ejs_model_GField_and_Potential_1D_v8wee.jar
https://dl.dropboxusercontent.com/u/44365627/lookangEJSworkspace/export/ejs_model_GFieldandPotential1Dv7EarthMoon.jar
https://dl.dropboxusercontent.com/u/44365627/lookangEJSworkspace/export/ejs_model_KeplerSystem3rdLaw09.jar
https://dl.dropboxusercontent.com/u/44365627/lookangEJSS/export/ejs_model_EarthAndSatelite.jar

## 1. Innovativeness of EJS

We have used the Easy Java Simulation (EJS) (Christian, Esquembre, & Barbato, 2011; Esquembre, 2012; Hwang & Esquembre, 2003) Authoring Toolkit and found it to be an innovative and versatile (generate java scripts models that runs on almost any mobile devices has a library of models free for anyone to change; allows for collaborative exploration sessions in Moodle, compliment remote laboratory usage; interfaces with sensors augmented reality and others) tool as it allows ordinary teachers to network with the Open Source Physics community. We present 4 computer models that the authors have customized to the Singapore syllabus for students' interactive learning. The authors have gained valuable experiences that sensitize us to effective teaching and learning with computers models such as augmenting with real equipment, where possible.

## 2. Rationale for use of computer models

We chose computer models as we believe physics is best learnt by inquiry through hands-on exploration. Many of our students fail to visualize the gravitational effects of planetary masses such as those causing the motions of satellites. We also saw word problem solving pedagogy to give students very little physical meaning.

## 3. The 5E Lesson design framework

The 5E lesson design framework is commonly used in the public schools and can be used to aid the inquiry-based lessons with computer models. By making use of practical situations for example the 15th February 2013 Chelyabinsk meteorite hitting Russia, students are more inclined to be *Engaged* in applying inquiry-based learning. Worksheets that detail the steps necessary to plan, experiment and measure to collect data encourage students to *Explore* the subject. *Explanation* of concepts can be done through student collaboration to share their evidence-based discussions. By questioning the students, the teacher is able to inspire *Elaboration* of content from students. Student presentations on the topic can *Evaluate* the given problem.

## 4. Computer models

Our computer models are available for download via the links and we welcome anyone to further customize them for their own teaching and learning needs, licensed Creative Commons Attribution. The simulations are updated to the latest EJS 5.0 beta version after feedbacks from students and teachers.

### *4.1. Gravity Mass Model*

Point Charge Electric Field in 1D Model by Andrew (Duffy, 2009), Professor of Physics, Boston University, USA served as the template for our Gravity Mass Model (Duffy & Wee, 2010a). The gravity mass model (Figure 1) is suited for the inquiry of gravitational field strength and potential of a 2 mass (M1 and M2) system with a test mass m, in the x-direction dimension. Invisible concepts of field strength and potential are modeled, allowing students to experience near 'impossible' isolated gravitational effects of small masses.

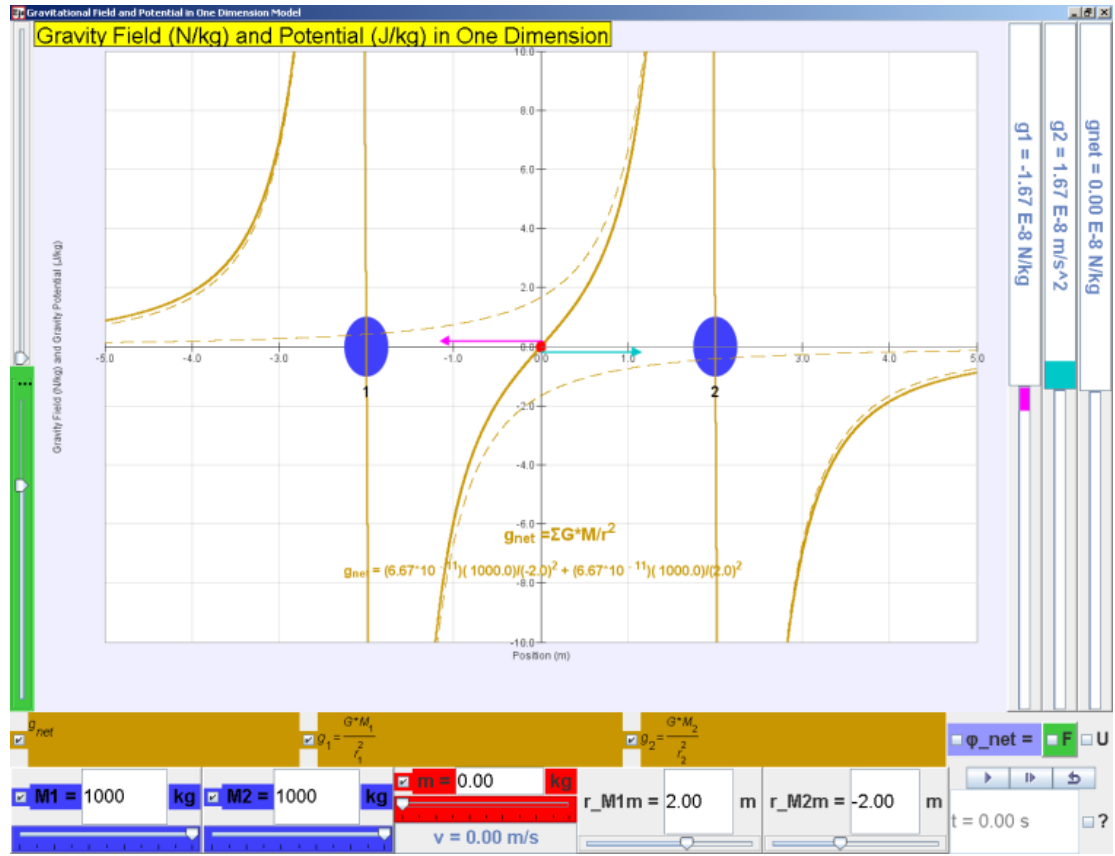

Figure 1. Gravity Mass Model(Duffy & Wee, 2010a) suitable for investigative inquiry learning through data collection, customized with syllabus learning objectives such as gravitational strength g, gravitational potential φ when one or both masses M1 and M2 are present with a test mass m. Superimpose are the mathematical representations, vector presentation of g, based on current Newtonian model of gravity.

### *4.2. Earth Moon Model*

The earth moon model (Figure 2) is suited for the inquiry of gravitational field strength and potential of the earth and moon system with a test mass m, in the x-direction. Concepts such as zero net force position and escape velocity of earth can be experienced and explored. These simulations provide simple means of doing practically impossible 'experiments' such as going to the Moon. We found the concepts of zero net force position and escape velocity to be especially useful in our classrooms with students. This is adapted from Duffy's (2009) work in Point Charge Electric Field in 1D Model which we customized to suit the gravitational concepts.

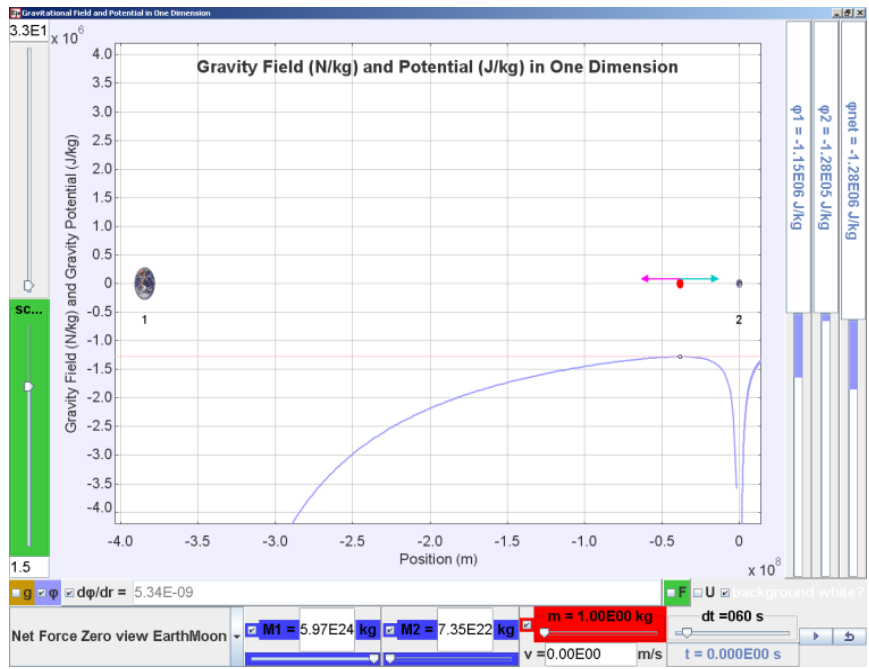

Figure 2. Earth Moon Model(Duffy & Wee, 2010b) suitable for investigative inquiry learning, further customized to allow the experiencing of an Advanced Level examination question June 87 /II/8. Data are based on real values where students can play and experience physics otherwise difficult to related to paper based question.

### 4.3. Geostationary Satellite Model

Earth and moon model by Francisco (Esquembre, 2010), Professor of Mathematics, University of Murcia, Spain, serves as the template for our geostationary satellite model (Figure 3). The geostationary satellite model (Wee & Esquembre, 2010) is suited for visualization of the three common orbits as well as some non-geostationary orbits. We even included incorrect physics such as a geostationary satellite above a point on the northern hemisphere of Earth. This is used to challenge thinking about what is 'wrong' with this orbit. A free body diagram showing the equal and opposite forces acting separately on the Earth and satellite helps students use what they learnt about Newton's Third Law in this context (Wee & Goh, 2013).

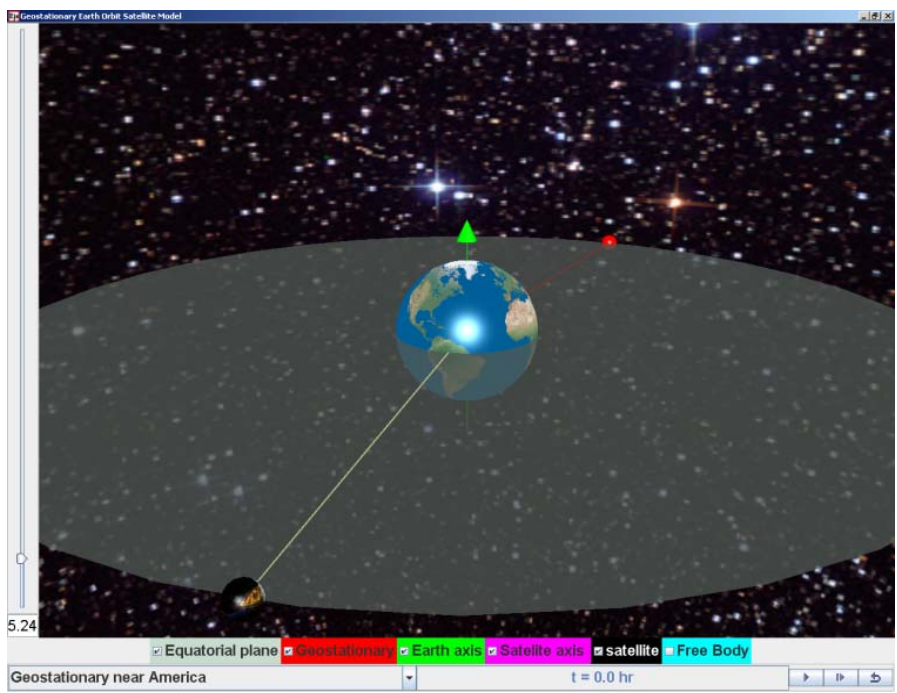

Figure 3. Geostationary Satellite around Earth Model (Wee & Esquembre, 2010) suitable for inquiry learning through menu selection. The geostationary checkbox option, 3D visualization, customized with

Singapore (red) and America (satellite) as a location position for satellite fixed about a position above the earth with 24 hours period, same rotation sense on the equator plane.

### 4.4. Kepler's System Model

The Kepler's Solar System Model (Timberlake, 2010), Professor of Physics, Berry College, USA serves as the template for our Kepler's System Model (Figure 4) (Timberlake & Wee, 2011). In addition to inquiry features-ideas in this paper by the same first author, (Wee, Goh, & Chew, 2013) we add the ability to use the EJS to collect and display the data instead of relying on pen and paper to record. This further supports our claim that EJS is an innovative tool for teacher professional development as the authors are able to customize and align the students' learning activity with the simulation tool. This would not be possible if the authors were to take existing simulations that cannot be customized.

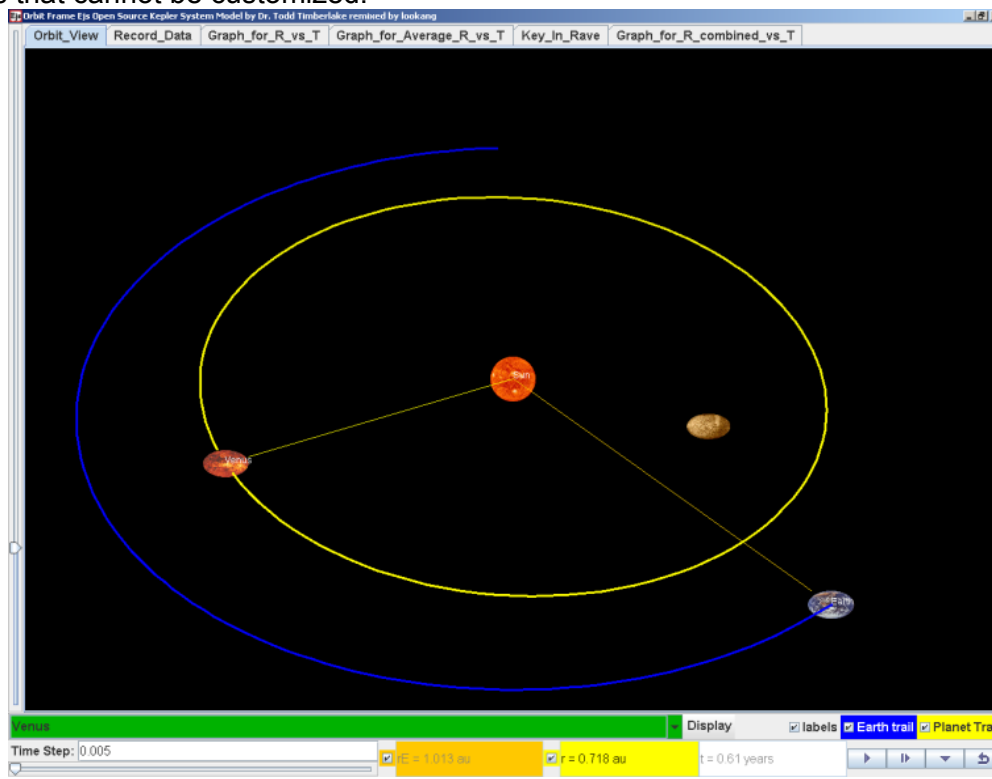

Figure 4. Kepler's System Model (Timberlake & Wee, 2011) with actual astronomical data built into the simulation, with realistic 3D visualization, (radius of planets such as Earth, rE and another planet for comparison r, and time t for determination of period of motion, T) data for inquiry learning and to situate understanding.

We have not tried out these newly developed features-ideas (Figure 5) for supporting learning but we speculate that students would approach the tasks now more like scientists, using computers to do routine work and seeing trends and patterns. EJS is also an effective tool for the authors because we are able to make EJS plot all the data points (Figure 6), with two possible data analysis to study trends of $T^2 = 0.999R^3$ (Figure 7) and $\lg T = 1.498 \lg R$ (Figure 8).

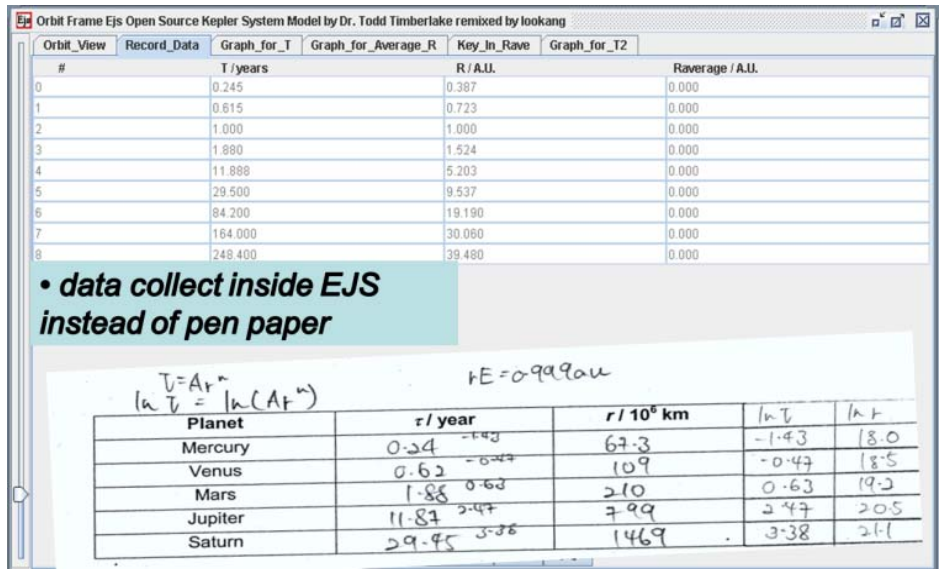

Figure 5.	Kepler's System Model (Timberlake & Wee, 2011) with tabs to Record Data showing (TOP) the data collectable by students after several runs of the model and (BOTTOM) sample pen paper recording of the data

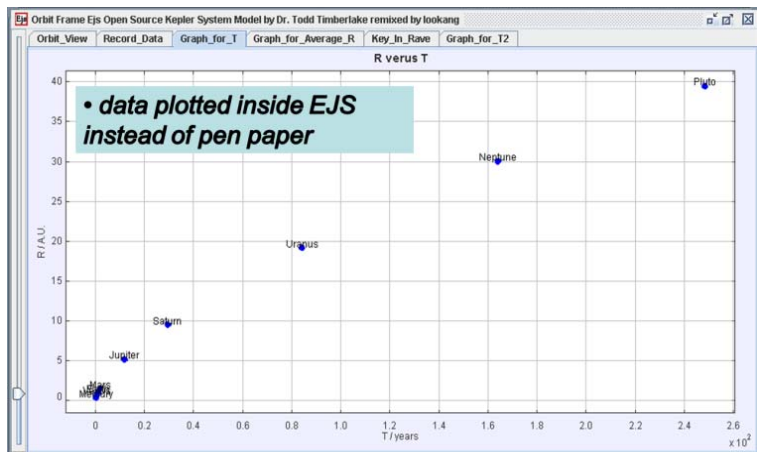

Figure 6.	Kepler's System Model (Timberlake & Wee, 2011) with tabs to show Graph for period T.

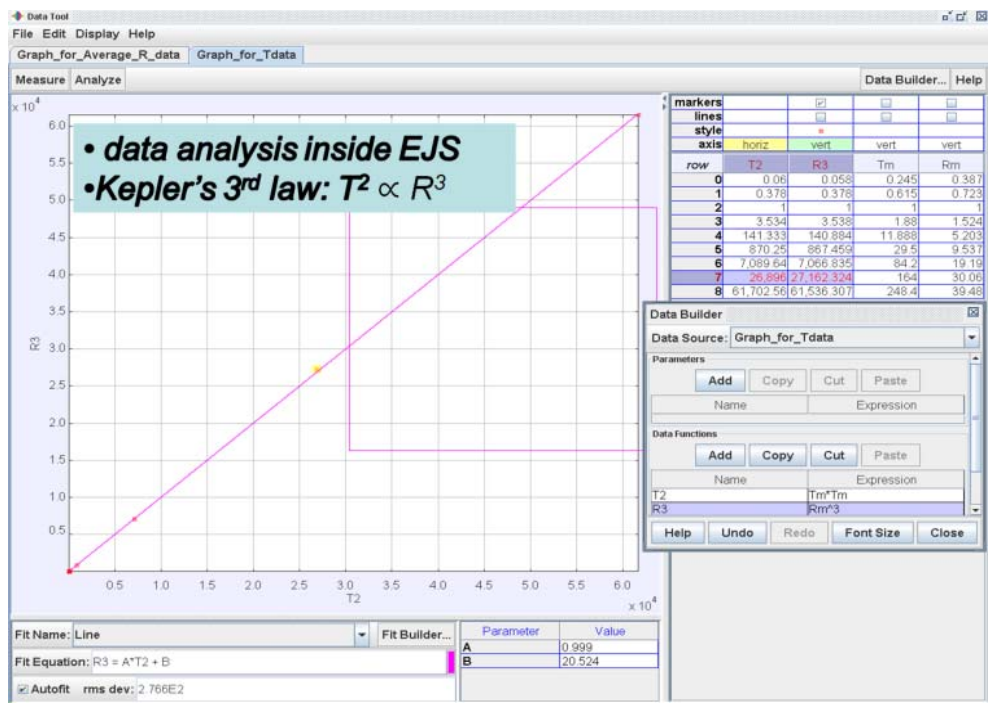

Figure 7.    Kepler's System Model (Timberlake & Wee, 2011) with data analysis tool for best fit line for $T^2 = R^3$ with coefficients A =0.999 suggesting to $T^2 \propto R^3$.

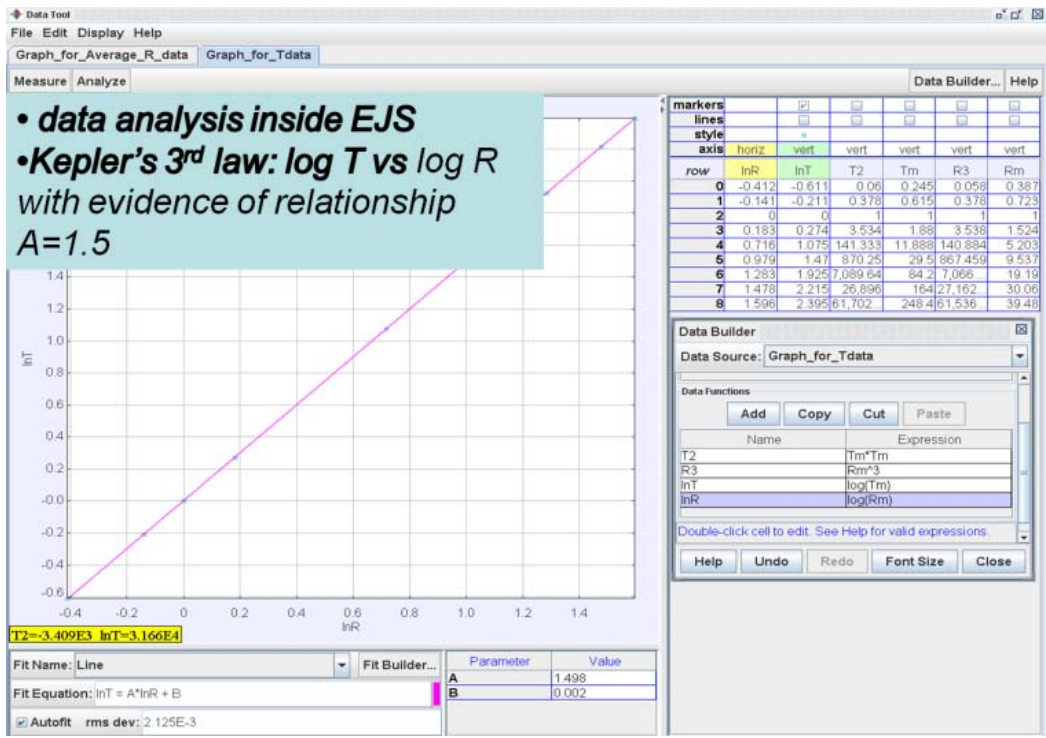

Figure 8.    Kepler's System Model (Timberlake & Wee, 2011) with data analysis tool for best fit line for $\lg T$ and $\lg R$ with coefficients A =1.498 suggesting $\lg T = 1.498 \lg R$.

We have briefly discussed the customized these four computer models for interactive engagement learning by standing on the shoulders of OSP giants. This is an innovative approach that deepens teacher's professional development in computer models design at practically zero dollars.

## 5. Conclusion

The authors have presented four gravity computer models that we have co-designed with teachers and students inputs, based on the earlier models-templates from the Open Source Physics (OSP) community. We argued for the use of EJS to design simulations as an innovative method for teacher professional development, and we advocate this professional learning for teachers who wish to customized simulations and eventually move on to teach students physics by coding-modeling in EJS.

The authors' school based research computer models, add credibility to this innovative approach of teachers' professional development through designing-customizing computer models that serves as effective inquiry tool that increases interactive engagement and visualization.

Feedback from the students has been positive, as triangulated from the survey responses, interviews with students and discussions with teachers.

The free EJS computer models can be downloaded from http://weelookang.blogspot.sg/2013/04/mptl18-easy-java-simulation-innovative.html, ComPadre Open Source Physics (Wee, 2012)and NTNU Virtual Physics Laboratory (Timberlake & Wee, 2011; Wee, Duffy, & Hwang, 2012a, 2012b; Wee & Esquembre, 2010) digital libraries. We hope more teachers around the world will further customize them for more intelligent (Juuti & Lavonen, 2006) guided inquiry in their own classes.

## 6. Acknowledgement

We wish to acknowledge the passionate contributions of Francisco Esquembre, Fu-Kwun Hwang and Wolfgang Christian for their ideas and insights in the co-creation of interactive simulation and curriculum materials.

This conference paper is made possible; thanks to the MOE-AST overseas training fund and eduLab project NRF2011-EDU001-EL001 Java Simulation Design for Teaching and Learning, awarded by the National Research Foundation, Singapore in collaboration with National Institute of Education, Singapore and the Ministry of Education (MOE), Singapore. Lastly, we also thank MOE for the recognition of our research and development on the computer models as a significant innovation with the 2012 MOE Innergy (HQ) GOLD Award.

**AUTHORS**

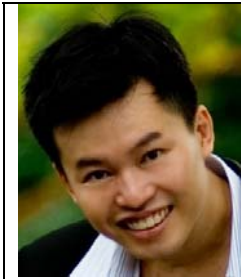

Loo Kang Lawrence WEE is currently an educational technology senior specialist at the Ministry of Education, Singapore. He was a junior college physics lecturer and his research interest is in Open Source Physics tools like Easy Java Simulation for designing computer models and the use of Tracker.

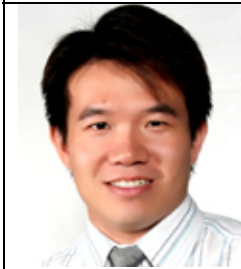

Giam Hwee Jimmy GOH is currently the Head of Science Department in Yishun Junior College, Singapore. He teaches Physics to both year 1 and 2 students at the college and advocates inquiry-based science teaching and learning through effective and efficient means.

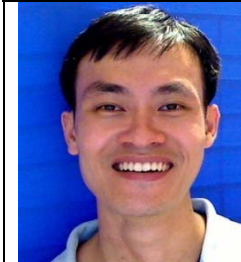

Ee Peow Stanley LIM is currently teaching in Anderson Junior College, Singapore. He is leading a Physics ICT Resource Team of teachers. Before that, he obtained a distinction in pre-service teaching practicum with his creative teaching methods.